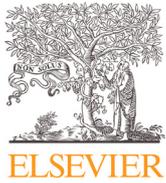
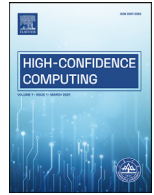

# Design high-confidence computers using trusted instructional set architecture and emulators

Shuangbao (Paul) Wang, Ph.D.

*Professor and Chair, Department of Computer Science, Morgan State University, Baltimore, MD, United States*

ABSTRACT

High-confidence computing relies on trusted instructional set architecture, sealed kernels, and secure operating systems. Cloud computing depends on trusted systems for virtualization tasks. Branch predictions and pipelines are essential in improving performance of a CPU/GPU. But Spectre and Meltdown make modern processors vulnerable to be exploited. Disabling the prediction and pipeline is definitely not a good solution. On the other hand, current software patches can only address non-essential issues around Meltdown. This paper introduces a holistic approach in trusted computer architecture design and emulation.

## 1. Introduction

High confidence computing needs trusted hardware and software. Virtualization is the cornerstone of cloud computing and provides performance, elasticity, and capability that otherwise could not be achieved on regular computing devices. There are many factors which could affect the security. At the software level, backdoors could exist in application programs that bring vulnerability to systems. At the compiler level, optimization could alter the normal functionality therefore violating the security. At the hardware level, Branch predictions and pipelines could leak the conditional or jump actions in a level 3 cache dump, where all virtual hosts share the same cache memory. At the chip level, manufactures could insert "logic bombs" in residual gates, understanding that not 100% of the gates are used in any processors on a single commonly "universal" chip.

This study focuses on security at the Instruction Set Architecture (ISA) level. Great attention is required in designing, developing, simulating, and emulating new processors that are immune to exploits and attacks.

Spectre and Meltdown make modern processors that use branch prediction and pipelines vulnerable to be exploited. It seems to be a virtualization problem, but if we look into the issues caused by Meltdown, it happens at the cache memory built inside the CPU, an essential memory that uses pipelines in improving performance of CPUs/GPUs. Disabling pipelines will lose the parallel speed-ups, therefore processors will slow down significantly. On the other hand, current software patches can only address non-essential issues around Meltdown. The goal for this research is to propose approaches in designing new generation processors that can be trusted and are immune to those attacks.

## 2. Related work

Von-Neumann architecture [1] is an implementation of Turing machine using stored-program concept. Turing Machine defines the theoretical concepts of computing mathematically and paths to solutions to these computations. The abstract computational device is defined as

$$T = \left(Q, \sum, s, \delta\right) \tag{1}$$

where: $Q$ is a finite set of states $q$, $\sum$ is a finite set of symbol, $s$ is the initial state $s \in Q$, and $\delta$ is a transition function determining the next move:

$$\delta : \left(Q \times \sum\right) \rightarrow \left(\sum \times \{L, R\} \times Q\right) \tag{2}$$

Instruction sets based on sequential access and linear computation have dominated the computer architecture for decades since 1945 until a reduced instruction set (all instructions run at same numbers of cycles) came to earth.

John L. Hennessy proposed the concept of reduced instructions set computing (RISC) in 1980s. The RISC architecture was focused on simpler, lower-cost microprocessors [2]. Since then, RISC architecture plays a major role in the design and development of CPUs. [3]. Recently, Oracle has launched Oracle Cloud Initiative that uses RISC chips to power its cloud platform. Apple has recently announced it will transition the Mac to its own custom silicon for leading performance and power AI/ML applications. The Apple New Silicon Initiative uses RISC architecture with improved security to reduce the risks of zero-day exploit and protect against remote code execution. Lacking security protection especially cache memory protection of existing processors has made systems vulnerable to side channel and cache timing attacks [4].

Spectre and Meltdown attacks [5,6] have once again raised concern about the trade-off between security and performance in architecture design [7]. Side channel related software leaks range from ALU leak, register bank leak, and debug leak. Mulder et al. proposed a modified architecture that seeds and masks circuit to protect side channel leak. The mask generation uses block ciphers, stream ciphers, hash functions, or other cryptographic primitives [8]. The approach has shown effective in protecting side channel attacks. The issue is the cryptographic computations are calculated on the same processor. Once the processor is compromised, the protection will be lost. Siddiqui et al. proposed a design to use FPGA over RISC-V implementation to prevent leakage over malware attacks. It establishes the relationships between BIOS and software during booting. The system uses TPM based boot to secure data flow within a processor [9]. The TPM based secure boot is effective to prevent leakage during the booting process. However, the data flow between the TPM and the processor could expose data to intruders.

QEMU on RISC-V has drawn the attention by industry and academia to be one of the top architecture and emulation platform due to their open architecture, cross platform, low latency, and optimized virtualization features [10,11]. Those advantages have made it especially useful in cloud computing environment. The trusted computing based QEMU virtual machine architecture improves the security for embedded devices [11].

This research proposes a holistic approach to enhance security and to seal the central processors from side channel, memory leak, cache dump, and debug leak attacks that could lead attackers to cross VMs in cloud computing environment. A quantum random number generator chip - Quantis QRNG is used to offer highest security, unbreakable and unpredictable randomness, and instant full entropy for cryptographic related computations.

## 3. Instruction set architecture

The Instruction Set Architecture determines the instructions, data types, addressing mode, and performance of a new processor. Karandikar [12] once said that "ISA is the most important interface in a computer system where software meets hardware". More effort has been put into improving the speed and performance of CPUs than that info security. With the recent discovery of vulnerabilities in CPUs, people use software patches to mitigate. This is apparently a short-sighted business decision. Essentially, it has to be resolved at the architecture level. The goal of this research is to discover a solution of designing new trusted ISA that can assure high-confidence computing.

### 3.1. Complex Instruction set computer architecture

John von Neumann proposed a stored-program computer architecture - von Neumann architecture - in 1945 [1]. The stored-program concept was based on universal Turing machine that was original proposed by Alan Turing in 1936 [13].

At ISA level, there are mainly two types of architecture: Complex Instruction Set Computers and Reduced Instruction Set Computers.

The Complex Instruction Set Computer (CISC) architecture uses the least amount of instructions needed for it to function. The hardware is capable of decoding and executing a series of different operations. For example, a $SHIFT$ instruction takes much less time than a complex $MULT$ instruction. The former may only need one clock cycle. The latter however may need a thousand or more clock cycles. This gives the compiler much less work to "translate" a high-level language statement into machine language (assembly).

### 3.2. Reduced instruction set computer architecture

The Reduced Instruction Set Computer (RISC) architecture only uses simple instructions that can be executed in registers within one clock cycle. Usually, RISC processors have large numbers of registers but few instructions. For the $MULT$ ($m \times n$) instruction, it divides into four steps: load number $m$ and $n$ into two registers A and B; $PROD$ A and B; and store the product to a memory location.

Patterson [14], Chen et al. [15] compared the similarities and differences between CISC and RISC, particularly the importance of pipelines of RISC architecture in today's computers.

RISC architecture emerged in IBM, CDC and other computers in 1960s. Now it has widely adoptions in microcontrollers, microcomputers, as well as mainframe computers.

It is hard to make a judgement whether RISC is better than CISC or not. But RISC has dominated the post-PC era and RISC-V [10] is becoming more popular in new processors, from micro-controller to super RISC chips [14].

### 3.3. Very long instruction word

Very Long Instruction Word (VLIW) computers divide a single VLIW instruction into multiple operations in execution. It is then possible to implement instructions with more than 100 bits on 32/64-bit processors.

### 3.4. Domain specific architectures

General purpose CPUs are dominant today. However, processors based on Domain Specific Architecture (DSA) are on the rise. Though Graphic Processing units (GPU) are being used as a general purpose processor, it was initially used in handling graphics for efficiency. Tensor Processing Units (TPU) handle tensor product [16], the most computational-intensive part of Deep Neural Network (DNN) inference in deep learning.

### 3.5. RISC-V

Developed at UC Berkeley in 2010, RISC-V (the fifth edition of RISC) is a new open ISA that contains a stack of software tools for architecture design. There have been a lot of hardware implementations such as $chisel-repo-tools$, a tool based on Python 3.7; $chipyard$, an agile RISC-V SoC design framework; $esp-isa-sim$, a RSIC-V ISA simulator; and $RocketChip$, and a parameterizable RISC-V chip generator, etc [10].

Many scholars agree that RISC is by far the best general-purpose ISA. As an open ISA, it reduces the risk of monopoly because it is unable to be owned by a single company, which may leave back-doors for vulnerabilities. It is also a good model for developing countries that utilize the tools to develop their own architectures for high confidence computing need.

Below are some major characteristics of RISC-V:

- Simplicity. The RISC-V ISA manual has 76,702 words (236 pages) while the x86-32 contains more than two million words (2198 pages).
- Modularity. The RISC-V core runs a full software stack from OS compilers to debuggers. This makes the multiply and divide, double-precision floating-point instructions possible.
- Efficiency. The RISC-V offers low energy implementation, which is ideal for IoT applications and also supports high-end applications
- Reservation of opcode space. Reservation of opcode space enables RISC-V to couple between general purpose cores and DSA cores. It also leaves opcode space for custom accelerators.
- Better Security. Implementations based on open source can be made for free, for profit, or closed for certain security needs. This eliminates the potential of malicious code or circuits in existing processors and allows it to be used in high confidence systems and applications.

DSA is becoming popular and general purpose ISA is still dominating the market. The new, trusted ISA should follow the RISC-V specification with the integration of security in design and development (both hardware and software) processes. Therefore, a better emulator that can emulate both hardware and software configuration is necessary.

## 4. Design and develop trusted instruction set architecture

The current ISAs focus more on performance by using pipelines, predictions, and shared multi-level cache memories. It lacks of guaranteed isolation between virtual machines in the cloud computing age. As a result, Spectre and Meltdown can happen on merely any processors manufactured after year 2000. So far software patches can only be effective if the processors running in a much slower mode with prediction and other critical speed related services disabled. This research proposes to seal the processors from side channel, memory leak and cache dump attacks at the ISA level. Quantum random number generator and trusted computing base add another level of security for cryptographic computations that are essential for data exchange and communication. This would have not made possible if the open RISC-V architecture were not come to earth.

Emulators help design and develop processor hardware such as CPU logic, I/O, and memory, and core software, such as kernels, OSs, and virtual machine managers. In cloud computing, processor resources are virtualized so vast numbers of services and instances can share the same resources. ISA emulators can not only assist in designing and developing a variety of ISA architecture such as x86-64, ARM, and SPARC but can also emulate virtualization modules, hypervisors, and dynamic binary translations. In addition, emulators such as Quick Emulator (QEMU) can emulate FPGA and help in reverse-engineering.

### 4.1. Hardware emulations

QEMU can emulates processors through dynamic binary translations. It is a full-system emulation system that can run operating systems for any machines on any supported software. It is especially helpful in designing and developing new processors that have better security mechanisms to prevent code embedding and potential logic bombs that come with an existing processors.

Below are some features of QEMU:

- Full ISA emulation
- User-model emulation
- Run virtual machines with near native performance
- Network emulation
- Guest isolation and TLS security

### 4.2. Vulnerabilities in virtualization

Virtualization is the core for cloud computing. Guest isolation is the fundamental part that all cloud platforms have claimed to provide. Yet, haven't been able to achieve due to the Spectre and Meltdown issues [6,16] discovered recently.

#### 4.2.1. Spectre
Modern processors use pipelines and branch prediction to maximize performance [17]. The CPUs guess the destination and attempt to fetch the instructions ahead for executions in order to keep the pipeline maximal. When there is a "hit", the CPU commits, if it is a "miss", the CPU discards. Spectre attacks look for the "miss" that leaks the victim's credentials and compromises the victim via a side channel. The vulnerability has been found on many common processors used in cloud and personal computing devices including Intel, AMD, and ARM.

Techniques to countermeasure the Spectre attack include preventing speculative execution, which results in significant decrease of CPU performance; modified speculation, which provides security but with reduced performance; and enhanced authentication and authorization in accessing each website in a separate process [18].

#### 4.2.2. Meltdown
Pipelines and out-of-order execution are indispensable in performance improvement for processors. Meltdown exploits side effects of out-of-order executions on modern processors. The attack is independent of the OS or other vulnerabilities in the software. The address space isolation break can lead to memory read out from other processes or virtualizations by an adversary.

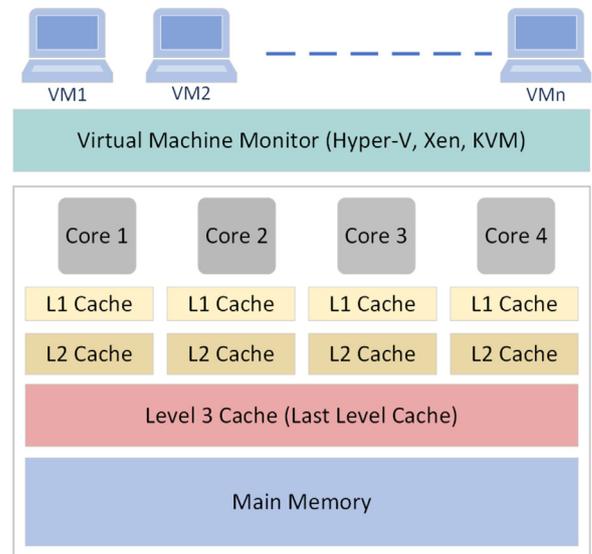

**Fig. 1.** Cache memory in cloud computing architecture.

Cache is faster memory than main memory. Modern CPUs have two to three levels of caches, each with larger space and less speed further away from the CPU. Cache functions as "whole-sale" where it only writes in blocks of data, instead of a single piece. Cache side-channel attacks exploit differences in memory access time. Flush + Reload [19] attacks exploit the last level cache (level-3) by frequently flushing a targeted memory location. The time differences in reloading the data give an indication of whether data was loaded by another process. CPUs and even cryptographic algorithms are both vulnerable to the attack. If a covert channel exists, it can leak information from one security zone to another. As a result, it is imperative to design trusted ISAs to prevent such attacks. Fig. 1 is a typical cloud architecture that the CPU with three levels of cache. Since level 3 cache is always shared between virtual hosts, a side channel attack could steal the cryptographic keys.

Software patches are common to countermeasure the Meltdown attack. Unfortunately, it brings significant overhead to processors, therefore reducing the speed. Hard memory segmentation is another way to address the problem. Intel and many processors have adopted memory segmentation technology. But existing kernel space and user space split is considered "soft".

### 4.3. Design and emulate trusted ISAs

The need to design and develop trusted ISAs for high confidence computing is growing. RISC-V has the support from leading industry including IBM, AMD, NVIDIA, Qualcomm, Microsoft, Google, and MIT. QEMU is the top pick to emulate a variety of processors and provides the best performance in a virtualized environment. The Capability Hardware Enhanced RISC Instructions (CHERI) extends the conventional ISAs with new feature for fine-grained memory protection and software compartmentalization [20]. The QEMU-CHERI combination on RISC-V provides hardware level security that can be emulated for real processors [21].

A discrete Trusted Platform Module (dTPM) is an isolated, separate feature chip that all necessary computing resources are contained within the discrete chip package. A discrete TPM has full control of dedicated internal resources including RAM, nonvolatile memory, and cryptographic logic.

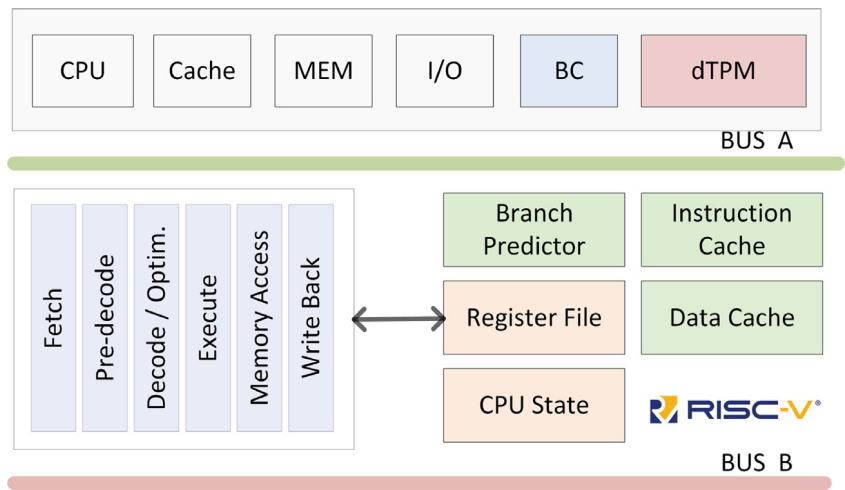

**Fig. 2.** Trusted ISA architecture.

This research uses a dual-bus model [22], discrete Trusted Platform Modules [23], the RISC-V architecture, and QEMU-CHERI hardware and software emulator to design and emulate a holistic Trusted ISA (TRISA) architecture that could be immune from Spectre and Meltdown attacks. Fig. 2 shows a new trusted ISA architecture. There are two buses in the proposed ISA. Bus A (green bus) is the internal bus between the green zone (top) and Demilitarized Zone (DMZ) (bottom). Bus B travels between the DMZ and the Internet. The DMZ (red zone) uses RISC-V architecture. Authentication and cryptographic key exchanges are handled by a dTPM, which contains a on-chip cryptographic engine for security.

*4.4. RISC-V chip development*

The constructing hardware in a Scala Embedded Language (Chisel) [24] is a hardware design language with advanced circuit generation features. In addition, it can be used for designing both ASIC and FPGA digital logic designs. The hardware primitives provides the power of modern programming language to developers to program complex and parameterizable circuits for a better level of abstraction and security.

*4.5. Enhanced RISC-V to side channel attacks*

The TRISA architecture applies EQMU-CHERI on top of RISC-V ISA architecture, a separate dTPM, a two-bus data flow, and debug feature to prevent side channel attacks. The protection model includes:

- *Lest privilege.* Setup bounds, permissions, and access control policies.
- *Strong Protection for Pointers.* Architectural primitives to support protection for C and C++ language pointers.
- *Use Trusted Computing Bases (TCB).* Protection of buffer overflow and leaks.
- *Integrity Protection.* Use hash and isolation control to guarantee the integrity.
- *Capacity Sealing.* Better exception control for modification and jump.
- *Instant full entropy.* Use Quantum RNG to offer highest security in cryptographic computations.

Mitigation on Spectre and Meltdown attacks can be implemented through:

- Branch avoidance on conditional moves.
- Use of speculation barriers to limit speculation.
- Avoid explicit flushes at the microarchitectural level.
- Use different kernel address spaces for different users.
- Integrity control at architectural level.

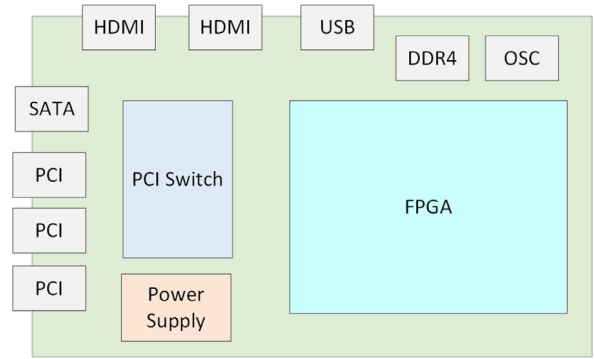

**Fig. 3.** Block diagram of microsemi HiFive unleashed expansion board.

These architectural changes and mitigations have to happen at a hardware level. It is imperative to consider these factors during the ISA design and emulation process so new processors can keep the pipeline speed-up performance while guarantee the principle of virtualization - isolation.

*4.6. Emulate TRISA on QEMU*

TRISA is being tested on a 64-bit Linux distribution (Ubuntu 18.10). After building QEMU with the RISC-V target, the system is ready to run RISC-V Linux on a development board - SiFive HiFive Unleashed or an Avalanche board with Microsemi PolarFire FPGA [25].

The HiFive board has universal features that can be tailored for development purpose. It contains:

- A low power PolarFire FPGA
- 24 lane PCIe switch and PCI express card connector
- SS and SATA connectors
- 4G DDR4 by 16 memory and flash memory
- FPGA design software

Fig. 3 shows a block diagram of Microsemi HiFive unleashed expansion board [26]. It is programmed for the ISA being tested.

**5. Conclusions and further discussions**

The 64-bit RISC-V TRISA emulation shows that through fundamental revision at the ISA level with considerations for keeping hardware speed-ups, the security is improved with the help of memory protection, better branch prediction control, preventing on-safe micro-architectural memory flush, enhanced integrity control, external crypto engine, and the use of the RISC-V TCB.

The holistic approach to enhance security at ISA level is a fundamental step to countermeasure Spectre and Meltdown attacks that otherwise would not be possible using software patches at the present level. The study brings the trusted computer systems to next generation high confidence computing that not only has implications in the science of computing but also has strong interests in national security.

**Declaration of Competing Interest**

The authors declare that they have no known competing financial interests or personal relationships that could have appeared to influence the work reported in this paper.

**Acknowledgment**

This research is funded in part by grants from NSF #2000136 and NSA #NCAEC-C-003-2020.